\makeatletter
\providecommand*{\input@path}{}
\g@addto@macro\input@path{{style/}}
\makeatother

\documentclass{article}
\usepackage[T1]{fontenc} 
\usepackage[utf8]{inputenc} 
\usepackage{ismir, amsmath, amssymb, cite, url, graphicx, color}

\usepackage{multirow}
\usepackage{booktabs}
\def\lbd{}	
\usepackage{subcaption}
\usepackage{lineno}
\usepackage{multicol}
\usepackage{array}
\newcolumntype{C}[1]{>{\centering\let\newline\\\arraybackslash\hspace{0pt}}m{#1}}

\title{Track Role Prediction of Single-Instrumental Sequences}

\threeauthors
  {Changheon Han} {Hanyang University \\ {\tt datajedi23@hanyang.ac.kr}}
  {Suhyun Lee} {Hanyang University \\ {\tt suhyeon00@gmail.com}}
  {Minsam Ko} {Hanyang University \\ {\tt minsam@hanyang.ac.kr}}

\def\authorname{Changheon Han, Suhyun Lee, and Minsam Ko}

\usepackage[bookmarks=false,pdfauthor={\authorname},pdfsubject={\papersubject},hidelinks]{hyperref}
\begin{document}

\maketitle
\begin{abstract}
In the composition process, selecting appropriate single-instrumental music sequences and assigning their track-role is an indispensable task. However, manually determining the track-role for a myriad of music samples can be time-consuming and labor-intensive. This study introduces a deep learning model designed to automatically predict the track-role of single-instrumental music sequences. Our evaluations show a prediction accuracy of 87\% in the symbolic domain and 84\% in the audio domain. The proposed track-role prediction methods hold promise for future applications in AI music generation and analysis.
\end{abstract}
\section{Introduction}\label{sec:introduction}

In the intricate process of musical creation, the determination of track-role in single-instrumental music sequences (e.g., main melody, sub melody, pad, riff, accompaniment, and bass) is undeniably paramount. Amidst a plethora of diverse sounds available to a composer, the act of selecting a specific sound and assigning it an appropriate role becomes a foundational determinant of the resulting musical piece. Instrumentation, a cardinal element in the craft of composition, also finds its articulate expression in track-role features, as suggested by previous studies~\cite{instrumentality04}.

Furthermore, the importance of track-role has been increasingly accentuated in the emerging discipline of controllable music composition employing deep learning. This avant-garde approach empowers composers and producers with unprecedented control over the creative process. By allowing the specification of instrument type, bpm, and measure count~\cite{mmm01} or even by inputting textual descriptions~\cite{musecoco02}, it offers a versatile canvas for musical experimentation. It is noteworthy that among the metadata incorporated for control, track-role has been identified as having a significant influence on the ultimate quality of the composed music~\cite{commu03}.

However, the task of sifting through myriad sounds and discerning their appropriate track-role is not without its challenges, often demanding considerable time and effort. Previous research has delved into the realm of audio samples, exploring avenues like identifying virtual instruments within a sample~\cite{showmeinst13} or classifying their inherent roles~\cite{audioroll07}. Such endeavors have been instrumental in assisting producers in pinpointing desired samples with enhanced efficiency. Building on this foundation, the present research introduces a novel approach leveraging deep learning techniques. Our methodology proposes an automatic discernment of track-role across both symbolic (e.g., MIDI) and audio domains, promising a more streamlined and effective music creation process.

\section{Proposed Methods}

This study introduces a deep learning model designed to predict the track-role of single-instrumental music sequences. For the purpose of classification, we consider a total of six target classes for track-role: Main Melody, Sub Melody, Pad, Riff, Accompaniment, and Bass. In terms of input data, both the symbolic domain and the audio domain were taken into consideration. Symbolic data, with its more explicit characteristics, offers potential for finer and more accurate distinctions. In contrast, the audio domain provides versatility, presenting applicability across a broader spectrum of scenarios. 

In our approach to each input domain, we employed a strategy of fine-tuning pre-trained models. First, for the symbolic domain data, we fine-tuned the MusicBERT model~\cite{musicbert08}. MusicBERT's initial training was executed on the expansive Million MIDI Dataset (MMD) which encompasses 1,524,557 songs. Meanwhile, a more compact variant, the MusicBERT small, was trained utilizing the Lakh MIDI Dataset (LMD), comprised of 148,403 songs. To adapt these models to our specific task, a classification and projection layer were appended to both. Subsequently, these enhanced models underwent a parameter update phase, utilizing our dedicated training data.

Transitioning to the audio domain, we integrated the PANNs model~\cite{panns09}, which was initially trained on a voluminous 5000 hours of AudioSet data~\cite{audioset10}. Furthermore, we also contemplated a variant of PANNs that incorporated attention feature fusion, referred to as PANN with AFF (w/aff)~\cite{pannaff11}. The audio inputs, sampled at a rate of 48 kHz, were transformed into log-mel spectrograms. For fine-tuning, two classification layers were added to the pre-trained PANNs, and training data updated the parameters. The attention feature fusion layer used initialized weights.

\renewcommand{\arraystretch}{1.4} 
\begin{table*}
    \small 
    \centering
    \begin{tabular}{cC{4.1em}C{4.1em}C{4.1em}C{4.1em}C{4.1em}C{4.1em}C{4.1em}C{4.1em}}
        \hline
        \multirow{2}{*}{} & \multicolumn{4}{c}{\textbf{Fine-Tuned}} & \multicolumn{4}{c}{\textbf{From-Scratch}} \\
        \cline{2-9}
         & \textbf{Accuracy} & \textbf{Precision} & \textbf{Recall} & \textbf{F1} & \textbf{Accuracy} & \textbf{Precision} & \textbf{Recall} & \textbf{F1} \\
        \hline
        MusicBERT\_base\textsuperscript{ \textasteriskcentered  } & \textbf{0.871} & 0.872 & 0.871 & 0.872 & 0.797 & 0.804 & 0.797 & 0.800 \\
        MusicBERT\_small\textsuperscript{ \textasteriskcentered  } & 0.853 & 0.851 & 0.853 & 0.852 & 0.783 & 0.781 & 0.783 & 0.782 \\
        PANNs w/o aff\textsuperscript{ \textdagger  } & 0.827 & 0.836 & 0.827 & 0.831 & 0.802 & 0.800 & 0.802 & 0.801 \\
        PANNs w/aff\textsuperscript{ \textdagger  }   & \textbf{0.843} & 0.850 & 0.843 & 0.846 & 0.820 & 0.820 & 0.820 & 0.820 \\
        \hline
    \end{tabular}
    \caption{Model Performance (\textsuperscript{ \textasteriskcentered  }: Symbolic domain model with ComMU, \textsuperscript{ \textdagger  }: Audio domain model with SCM)}
    
    \label{tab:model_performance}
\end{table*}

\section{Experimental Setup}\label{subsec:body}

\subsection{Datasets}
We utilized the ComMU dataset, containing 11,144 MIDI samples, for training and evaluation. To balance sample distribution across track roles, we selected 500 samples from each role, based on the smallest class, "base." Given ComMU's symbolic nature, we developed the Synthesized-ComMU (SCM) dataset for the audio domain. This was achieved by aligning ComMU's instrument data with the NSynth dataset~\cite{nsynth12} and selecting randomized instrument presets for consistent timbre. The MIDI rendering pipeline{\footnote{https://github.com/spear011/NSynth-MIDI-Renderer-for-massive-MIDI-dataset}} and SCM{\footnote{https://github.com/spear011/SCM-Dataset}} can be accessed online.

\subsection{Training Details}
We partitioned the ComMU and SCM datasets at a ratio of 8:2 for model training and evaluation, respectively. Additionally, 10\% of the training data was earmarked as a validation set to monitor the model's performance during the training phase. Data augmentation by manipulating BPM and audio key were applied to the training dataset to ensure model robustness. For the optimization process, the Adam optimizer was selected with a learning rate that initially peaked at 5e-5 and was gradually reduced over the course of training. When fine-tuning the pre-trained MusicBERT, we employed a step size of 8100, amounting to 4 epochs. In contrast, training from scratch involved a larger step size of 40500, culminating in 20 epochs. As for the PANNs models, irrespective of whether attention feature fusion was incorporated, a consistent step size of 20250 was used, which corresponded to 10 epochs.

\begin{figure}[t]
    \begin{subfigure}[b]{0.225\textwidth}
        \centering
        \includegraphics[width=\textwidth]{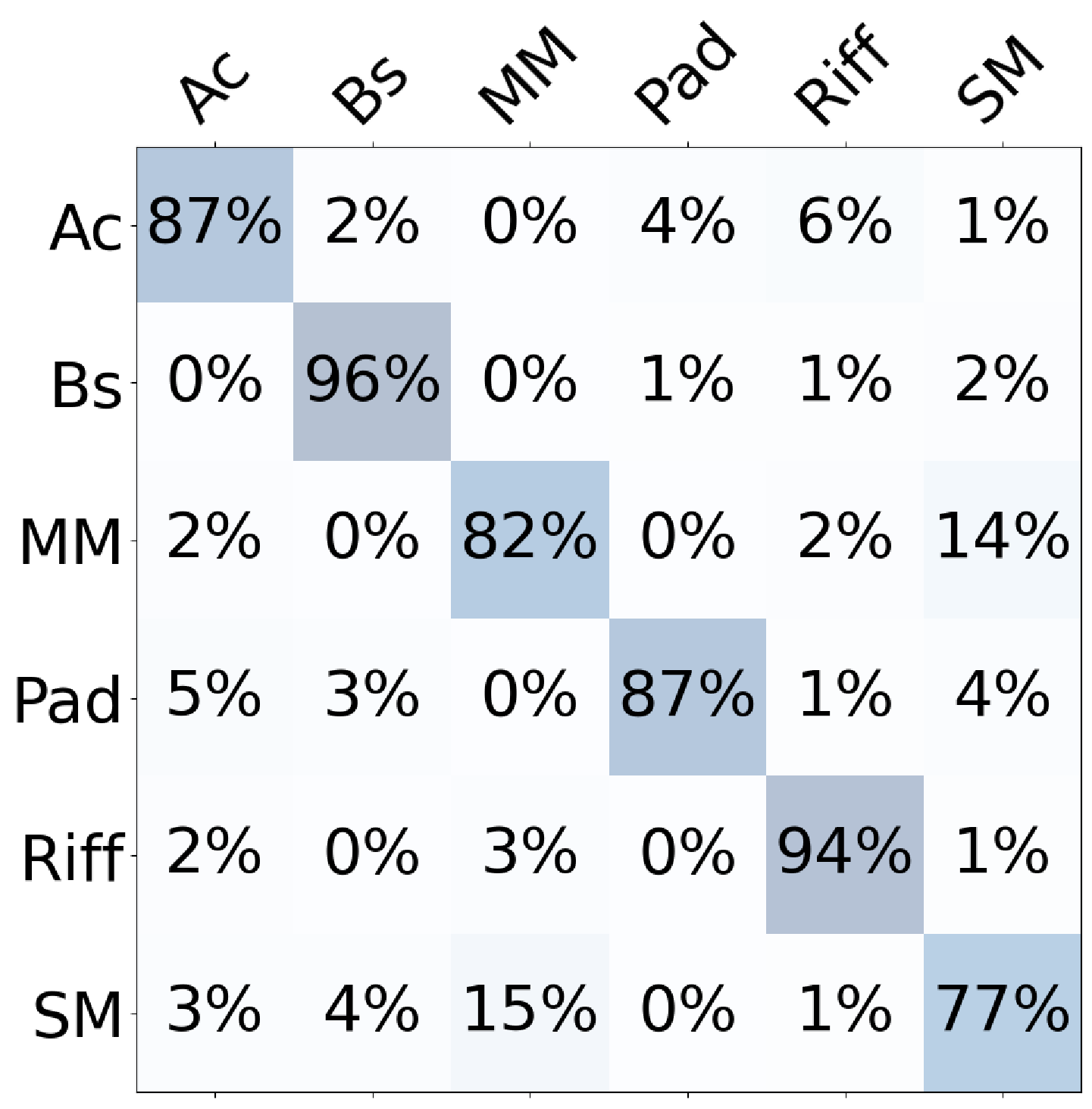}
        \caption{MusicBERT base}
        \label{fig:subfig1}
    \end{subfigure}
    \hfill
    \begin{subfigure}[b]{0.225\textwidth} 
        \centering
        \includegraphics[width=\textwidth]{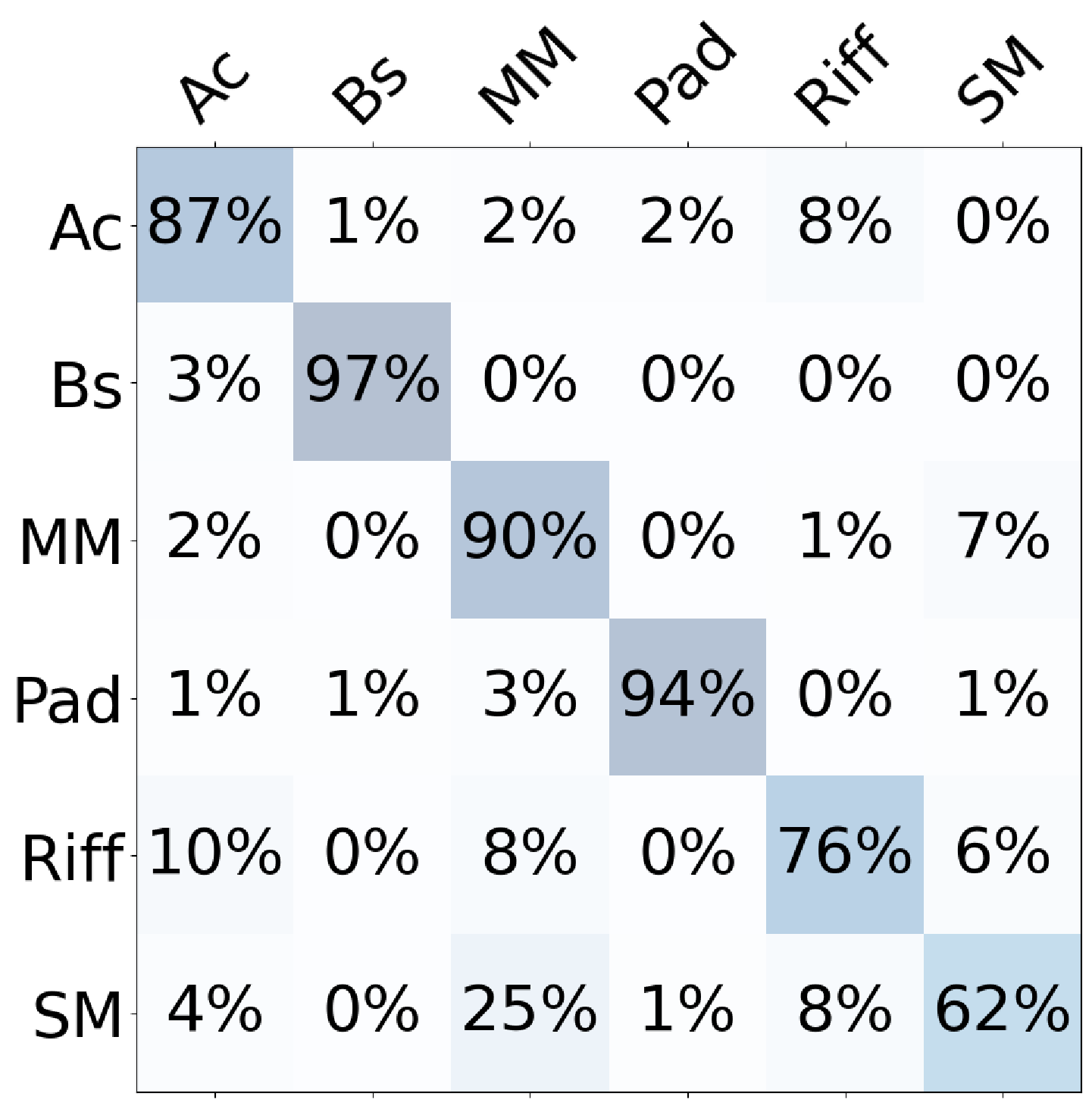}
        \caption{PANNs w/aff}
        \label{fig:subfig2}
    \end{subfigure}
    \caption{Model Confusion Matrix (Ac: Accompaniment, Bs: Bass, MM: Main Melody, SM: Sub Melody)}

    \label{fig:confusion}
    \vspace{-1em}    
\end{figure}

\section{Results}\label{sec:page_size}

Table \ref{tab:model_performance} shows the performance results of each approach. Notably, in both the Symbolic and Audio domains, models that employed the fine-tuning strategy on pre-trained models consistently outperformed those trained from scratch with identical architectures. The models in the Symbolic domain that used MIDI data as input demonstrated superior performance compared to their Audio domain counterparts. However, the difference in performance was not vast. The best-performing model for the Symbolic domain was the fine-tuned MusicBERT Base, achieving an accuracy of 0.871. In the Audio domain, the pinnacle of performance was reached by the fine-tuned PANN model with attention fusion, recording an accuracy of 0.843.

Figure \ref{fig:confusion} illustrates the class-wise accuracy of the top-performing models. A consistent trend was observed across both domains. Both models tended to struggle particularly when discerning between the Main Melody and Sub Melody. For instance, a simplistic melodic contour characterized by short measures interspersed with elongated notes was erroneously predicted as a Sub Melody by the model, even though the correct label was Main Melody. Further, while analyzing the Audio domain model, instances were observed where the Riff class was mistakenly predicted as Accompaniment or Main Melody. To illustrate, a sequence comprising short notes in a repetitive pattern was designated as a Riff by the model, but the ground truth labeled it as Accompaniment.

\section{Conclusion}
This study introduces a method to predict the track-role of music sequences in both the Symbolic and Audio domains, utilizing fine-tuned deep learning models. The automatically predicted Track-role data holds promise for future applications, including efficient sample search and management, as well as advancements in AI-assisted composition research. Experimental findings highlighted certain track-role classes where prediction performance was suboptimal, often corresponding to sequences exhibiting diverse musical forms. To bolster future performance, a systematic approach addressing the vast spectrum of musical structures becomes imperative. One potential avenue could be adopting a learning strategy like \textit{curriculum learning}, which incrementally tackles data that poses distinction challenges, presenting a promising direction for enhancement.

 \onecolumn \begin{multicols}{2}
 
\section{Acknowledgments}
This work was supported by Institute of Information \& communications Technology Planning 
\& Evaluation (IITP) grant funded by the Korea government (MSIT) (No.RS-2022-00155885, Artificial Intelligence Convergence Innovation Human Resources Development (Hanyang University ERICA)).

\bibliography{main}

 \end{multicols}

\end{document}